\documentclass{emulateapj}
\usepackage{apjfonts}
\usepackage{amsmath}
\def\j{{\scriptscriptstyle (j)}}
\def\tot{{{\rm tot}}}
\def\sub{{{\rm sub}}}
\def\obs{{{\rm obs}}}
\def\dep{{{\rm dep}}}

\shorttitle{Bimodal distribution of GRBs}
\shortauthors{Toma, Yamazaki, \& Nakamura}
\begin{document}

\title{A Possible Origin of Bimodal Distribution of Gamma-Ray Bursts}
\author{Kenji Toma$^{1}$, Ryo Yamazaki$^{2}$
and Takashi Nakamura$^{1}$}
\affil{
$^{1}$Department of Physics, Kyoto University,
Kyoto 606-8502, Japan
\\
$^{2}$Department of Earth and Space Science,
Osaka University, Toyonaka 560-0043, Japan
}

\email{toma@tap.scphys.kyoto-u.ac.jp}

\begin{abstract}
We study the distribution of the durations of gamma-ray bursts (GRBs)
in the unified model of short and long GRBs 
recently proposed by Yamazaki, Ioka, and Nakamura. 
Monte Carlo simulations show clear bimodal distributions,  
with lognormal-like shapes for both short and long GRBs,
in a power-law as well as a Gaussian angular distribution of the
subjets. 
We find that the bimodality comes from the existence of the discrete
emission regions (subjets or patchy shells) in the GRB jet. 
To explain other temporal properties of short and long GRBs, the subjet
parameters should depend on the angle in the whole jet. 
\end{abstract}



\section{Introduction}
The durations of gamma-ray bursts (GRBs) observed by BATSE
show a bimodal distribution,
which has led to a classification of GRBs into two groups:
bursts with $T_{90}$ durations $<2~\rm{s}$ are called short GRBs, and
those with durations $>2~\rm{s}$ are called long GRBs \citep{k93,m94}.
If $T_{90}$ directly reflects the active time of the progenitor of the
GRB, different origins of short and long bursts are implied, such that
the former arise from the binary neutron star mergers
while the latter arise from the collapse of massive stars
\citep[e.g.][]{m02,zm04}.

The short and long bursts roughly consist of 25\% and 75\%,
respectively, of the total BATSE GRB population.
We should regard these fractions as comparable,
considering possible instrumental effects on the statistics.
If these two phenomena arise from essentially different origins,
the similar number of events is just by chance.
However, some observations have suggested
that the short GRBs are similar to the long GRBs
\citep[e.g.,][]{germany00,lrg01,nakar02,lamb2003b,ggc04}.
Motivated by these facts, \citet{yin04b}
proposed a unified model of short and long GRBs,
even including X-ray flashes (XRFs) and X-ray--rich GRBs,
and showed that it is possible to attribute
the apparent differences of the light curves and spectra
of these four kinds of events
to the different viewing angles of the same GRB jet.
This is a counter-argument against the current standard scenario of the
origins of short and long GRBs.

In this paper, we perform Monte Carlo simulations to show
that our unified model naturally leads
the bimodal distribution of the $T_{90}$ durations of GRBs.
The paper is organized as follows.
In \S~\ref{sec:model} we begin with a brief review of
our unified model of short and long GRBs.
The $T_{90}$ duration distribution is
calculated in \S~\ref{sec:duration}.
Section \ref{sec:discuss} is devoted to discussions.

\section{Unified Model of Short and Long GRBs}
\label{sec:model}
We briefly describe our unified model of short and long GRBs
\citep[for details, see][]{yin04b}.
We assume that
the GRB jet is not uniform but made up of multiple subjets,
and that each subjet causes a spike in the observed light curve.
This is an extreme case of an inhomogeneous jet model
\citep{nakamura2000,kp00}.
Let us consider a subjet with the opening half-angle $\Delta\theta_\sub$
moving with Lorentz factor $\gamma$,
observed from the viewing angle $\theta_v$.
Because of relativistic effects,
the subjet emission becomes dim and soft
when $\theta_v$ is larger than $\sim \Delta\theta_\sub+\gamma^{-1}$
\citep{in01}.
The {\it effective} angular size of its emission region is
$\pi(\Delta\theta_\sub+\gamma^{-1})^2$,
which is larger than the geometrical size of $\pi{\Delta\theta_\sub}^2$.
For the multiple subjet case, the crucial parameter is
the multiplicity ($n_s$) of the {\it effective} emission regions
along a line of sight .
If many subjets point toward us (i.e., $n_s\gg1$) the event looks like a
long GRB,
while if a single subjet points toward us (i.e., $n_s=1$) the event
looks like a short GRB.

\begin{figure}
\plotone{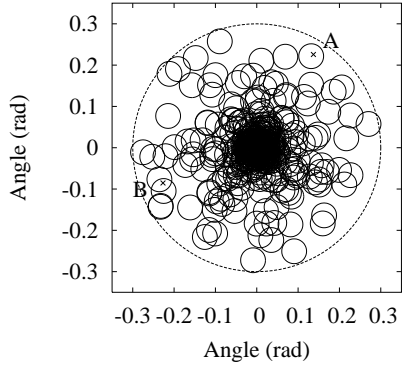}
\caption{
Angular distribution of $N_\tot=350$ subjets confined
in the whole GRB jet in our simulation.
Each subjet is located according to the power-law distribution function of
eq.(\ref{eq:jetP}).
The whole jet has an opening half-angle of
$\Delta\theta_\tot=0.3$~rad.
The subjets have the same intrinsic luminosity and
opening half-angles $\Delta\theta_\sub=0.02$~rad, and the
other properties are $\gamma=100$, $r=3 \times 10^{13}$\,cm,
$\alpha_B=-1$\,, $\beta_B=-2.5$\, and $\gamma h{\nu'}_0=500$\,keV.
The effective angular size of the subjets 
are represented by the solid circles, while 
the whole jet is represented by the dashed circles.
The examples of lines of sight A and B are shown in
the figure, while
C is located at $(-0.04~\rm{rad}, 0.04~\rm{rad})$
and D is close to the center of the whole jet.
}\label{jetP}
\end{figure}

Below we give a typical set of parameters
for the temporal and spatial configurations of the GRB jet
to demonstrate which type of event is observed depending on $n_s$.
We suppose that $N_\tot$ subjets are launched
from the central engine of the GRB randomly in time and directions and 
that the whole jet consists of these subjets.
We introduce a spherical coordinate system $(r, \vartheta, \varphi)$
in the central engine frame,
where the origin is the location of the central engine, and
$\vartheta=0$ is the axis of the whole jet.
The axis of the $j$th subjet ($j=1,\,\cdots,\,N_\tot$) is denoted
by $(\vartheta^\j, \varphi^\j)$,
while the direction of the observer is denoted by
$(\vartheta_\obs, \varphi_\obs)$.
We suppose that the $j$th subjet departs at time $t_\dep^\j$
from the central engine
and emits at radius $r=r^\j$ and time
$t=t_\dep^\j+r^\j/\beta^\j c$.
The departure time of each subjet $t_\dep^\j$
is assumed to be homogeneously random
between $t=0$ and $t=t_{\rm dur}$,
where $t_{\rm dur}$ is the active time of the central engine measured in
its own frame and is set to $t_{\rm dur}=20$~s.
The emission model for each subjet is the same
as the uniform jet model in \citet{yin03b}.
For simplicity, all the subjets are assumed to have the same
intrinsic luminosity and opening half-angle
$\Delta\theta_\sub^\j=0.02$~rad,
and the other properties are $\gamma^\j=100$,
$r^\j=3\times10^{13}$~cm,
$\alpha_B^\j=-1$, $\beta_B^\j=-2.5$,
and $\gamma h{\nu'}_0^\j=500$~keV for all $j$.
The opening half-angle of the whole jet is set to
$\Delta\theta_\tot=0.3$~rad.
We randomly spread $N_{\tot}=350$ subjets following
the angular distribution function of the subjets as
\begin{equation}
\frac{dN}{d\Omega} 
\equiv n(\vartheta,\varphi) = \begin{cases}
n_c, & 0<\vartheta<\vartheta_c, \\
n_c (\vartheta/\vartheta_c)^{-2},
& \vartheta_c<\vartheta<\vartheta_b,
\end{cases}
\label{eq:jetP}
\end{equation}
where $\vartheta_b=\Delta\theta_\tot-\Delta\theta_\sub$ and
$\vartheta_c=0.02$~rad \citep[see also][]{rlr02,zm02}.
Figure~\ref{jetP} shows an example of the angular distribution of the
effective emission regions of the subjets in our calculation.
Most of the subjets are concentrated near the $\vartheta=0$ axis
(i.e., the multiplicity in the center $n_s\sim 100$).
For our adopted parameters,
isolated subjets exist near the edge of the whole jet,
and there are some directions in which no subjet is launched.

\begin{figure}
\plotone{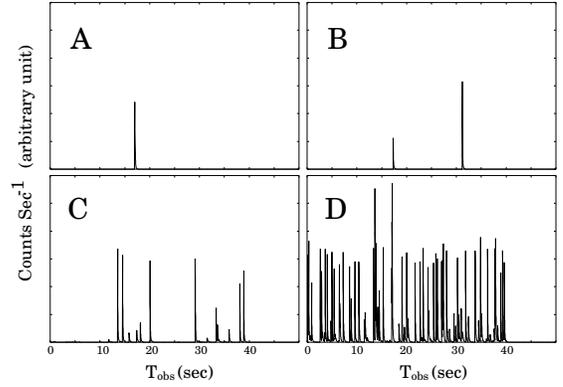}
\caption{
Observed light curves in the 50--300~keV band for the lines of sight
shown in Fig.~\ref{jetP}:
A with $n_s=1$ ({\it upper left}), B with $n_s=2$ ({\it upper right}),
C with $n_s=15$ ({\it lower left}),
and D with $n_s=97$ ({\it lower right}).
The sources are located at $z=1$.
The $T_{90}$ durations are $0.25$~s for A, $14.1$~s for B,
$25.4$~s for C, and $37.8$~s for D.
}\label{lcs}
\end{figure}

Figure~\ref{lcs} shows examples of the observed light curves
in the 50--300~keV band,
each of which corresponds to the lines of sight A, B, C, and D in
Figure~\ref{jetP}. 
The coordinate $(\vartheta_\obs, \varphi_\obs)$ of C is 
$(-0.04~\rm{rad}, 0.04~\rm{rad})$, and
D is close to the center of the whole jet.
If many subjets point in the direction of the line of sight,
such as in the cases of C ($n_s=15$) and D ($n_s=97$), we see a spiky
temporal structure.
In the case of B ($n_s=2$), the event consists of
the distinct emission episodes.
These are identified as long GRBs.
If only one subjet points toward us,
like in the case of A ($n_s=1$),
the contributions to the observed light curve from the other subjets
are negligible because of relativistic beaming effect,
so that the observed gamma-ray fluence and duration are both about a
hundredth of the typical values of long GRBs.
These are quite similar to the characteristics of short GRBs.
In addition,
when the line of sight is away from any effective subjet regions
(i.e., $n_s=0$),
the soft and dim prompt emission is observed because of relativistic Doppler
effect and beaming effect,
which is identified as an XRF or an X-ray--rich GRB
\citep{in01,yin02,yin03b,yin04a,yin04b,yyn03}.

\begin{figure}
\plotone{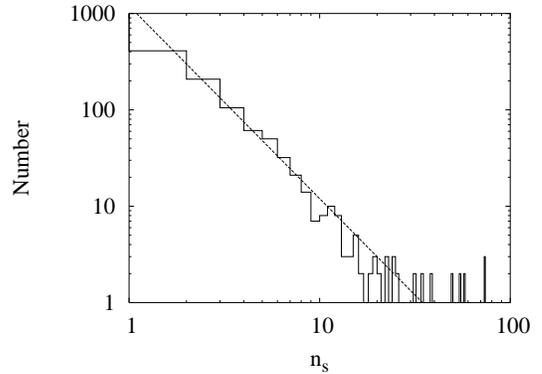}
\caption{
Distribution of multiplicity $n_s$
for the angular distribution of the subjets of Fig.~\ref{jetP}.
The dashed line represents the analytical estimate of the
${n_s}^{-2}$ line (see text).
}\label{multiP}
\end{figure}

\section{Distribution of $T_{90}$ Duration}
\label{sec:duration}

We perform Monte Carlo simulations
to show that our unified model can explain
the observed bimodal distribution of $T_{90}$ durations of GRBs.
We fix the subjets' configuration as in Figure~\ref{jetP}.
We vary only the line of sight of the observer and
calculate the $T_{90}$ duration for each observer in the 50--300~keV band.
We generate 2000 lines of sight with $0 < \vartheta_\obs < 0.35$~rad
according to the probability distribution of
$\sin\vartheta_\obs\,d\vartheta_\obs\,d\varphi_\obs$.
We then select only hard events, whose observed hardness ratio is
$S(2-30~{\rm keV})/S(30-400~{\rm keV}) < 10^{-0.5}$
\citep{s03}.
The other soft events are classified as XRFs or X-ray--rich GRBs,
which are observed when all subjets are viewed off-axis.

\begin{figure}
\plotone{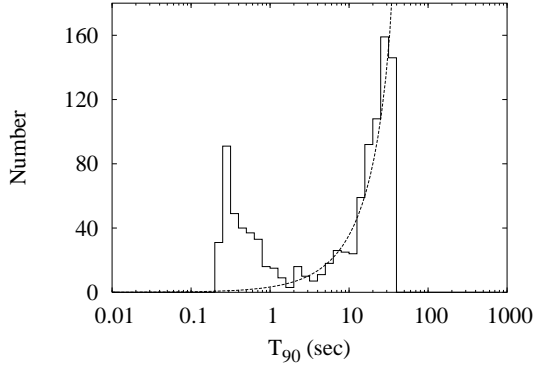}
\caption{
$T_{90}$ duration distribution in the 50--300~keV band of hard events
 with observed 
fluence ratio $S(2-30~{\rm keV})/S(30-400~{\rm keV}) < 10^{-0.5}$.
The jet model is the power-law.
All sources are located at $z=1$.
The dashed line represents the analytical formula for the long GRBs,
given by eq.~(\ref{eq:appP}).
}
\label{dgP}
\end{figure}

Figure \ref{multiP} shows the distribution of $n_s$ in our simulation.
The multiplicity $n_s$ is roughly
proportional to $n(\vartheta_\obs, \varphi_\obs)$.
Then the distribution of $n_s$
is given by
$P(n_s) \propto \sin(\vartheta_\obs)(d\vartheta_\obs / dn_s) \sim
n_s^{-2}$ (Fig.~\ref{multiP}, {\it dashed line}).
We first consider the $T_{90}$ distribution in the case in which the
redshifts of all the sources are fixed at $z=1$ for simplicity.
The result is shown in Figure~\ref{dgP}.
One can see a bimodal distribution of $T_{90}$ clearly.
Which type of burst is observed, long or short, depends on $n_s$,
and the distribution of $n_s$ is unimodal.
Then why does the distribution of the duration become bimodal?
The reason for the scarcity of the events for $1 < T_{90} < 10$~s is as
follows. 
Let us first consider the event with $n_s=1$.
In this case the $T_{90}$ duration does not vary significantly
around $\sim 0.25$~s when $\theta_v<\Delta\theta_\sub$,
which is determined by the angular spreading time of a subjet.
As the viewing angle increases, $T_{90}$ increases \citep{in01}.
When $\theta_v\gtrsim\Delta\theta_\sub+\gamma^{-1}$, however,
the emission becomes soft and dim, 
so that the event will not be detected as a GRB \citep{yin02,yin03b,yyn03}.
The $T_{90}$ takes a maximum value of $\sim 0.75$~s when
$\theta_v\sim\Delta\theta_\sub+\gamma^{-1}$.
We confirm that $n_s=1$ for almost all $T_{90} < 1$~s events.
Next let us consider the $n_s=2$ case.
The example of the light curve for this case is Figure~\ref{lcs}$b$,
and the $T_{90}$ is 14.1~s.
The $T_{90}$ duration is roughly given by
the interval between the arrival times of two pulses.
Since the two pulses arrive sometime in the range $0<T_\obs<T_{\rm dur}$,
where $T_{\rm dur}$ is the active time of the central engine
measured in the observer's frame,
$T_{\rm dur}=(1+z)t_{\rm dur}=40$~s,
the mean interval is 40/3=13.3~s. 
This means that the duration of the $n_s=2$ event is much longer than
that for $n_s=1$. 
For $n_s \geq 3$, the
mean duration is longer than 13.3~s.
The typical example is Figure~\ref{lcs}$c$ for $n_s=15$,
with $T_{90}=25.4$~s.
This is the reason we have few events for
$1 < T_{90} < 10$~s.
The maximum value of $T_{90}$ is $\sim T_{\rm dur}$.
For the long bursts, the distribution function of $T_{90}$ durations
can be derived from a simple probability argument
(see the Appendix~\ref{sec:app} for details).
The dashed line in Figure~\ref{dgP} represents the analytical
formula of equation (\ref{eq:appP}).
On the other hand, the distribution function of the short bursts seems
to be too complicated to calculate analytically,
since it sensitively depends on the jet configurations,
such as the angular distribution and the intrinsic properties of the
subjets.

\begin{figure}
\plotone{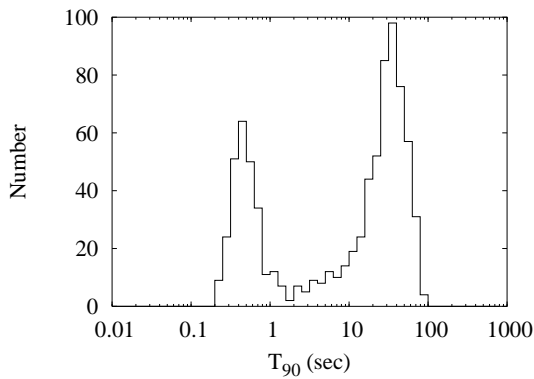}
\caption{Same as Fig.\ref{dgP} but the source redshifts are varied
according to the cosmic star formation rate(see text for details). Both
short and long GRBs look like lognormal distributions.}
\label{dgzP}
\end{figure}

The ratio of events of the short GRBs
and the long GRBs is about $2:5$, which can be
explained as follows \citep{yin04b}.
The event rate of the long GRBs is in proportion to
the effective angular size of the central core
$\vartheta_{c,eff}{}^2\sim(0.15~{\rm rad})^2$, where $n_s\geq2$.
The event rate of the short GRBs is in proportion to
$M(\Delta\theta_\sub+\gamma^{-1})^2$, where
$M$ is the number of isolated subjets in the
envelope of the core and $M\sim10$ in our present case.
Then the ratio of event rates of the short and long GRBs becomes
$M(\Delta\theta_\sub+\gamma^{-1})^2:{\vartheta_{c,eff}}^2\sim2:5$.

In reality, we should take into account the source redshift distribution.
We assume that the rate of GRBs is in proportion to
the cosmic star formation rate.
We adopt the model SF2 in \citet{porciani2001}, in which
we take the standard cosmological parameters of
$\Omega_M =0.3$ and $ \Omega_{\Lambda}=0.7$.
Figure~\ref{dgzP} shows the result. The distribution is again clearly
bimodal, and the shapes of the short and long GRBs look like lognormal
distributions. 
The ratio of the number of short and long GRBs is about $2:5$ in
this case as well.
The dispersion of the lognormal-like distribution seems relatively small
compared to the observations.
This is ascribed to simple modeling in this paper.
We fix the jet configuration and
use the same intrinsic properties of the subjets.
If we vary $t_{\rm dur}$ for each source and
$\gamma^{\j}$ for each subjet randomly, for example,
the dispersion of lognormal-like $T_{90}$ duration distribution
will increase from the general argument that the dispersion of the
lognormal
distribution increases with the increase of the number of the associated
random variables \citep{in02}.
In more realistic modeling
the observed dispersion will be reproduced.

\begin{figure}
\plotone{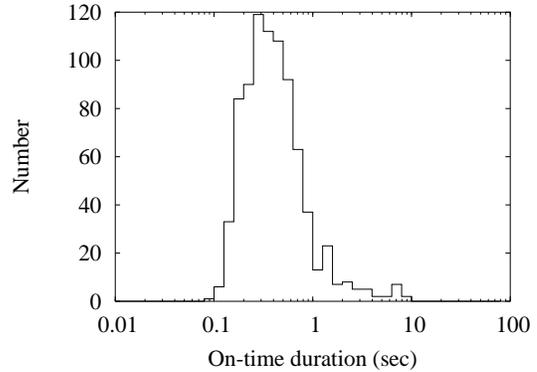}
\caption{
{\scriptsize ON} time duration distribution in the 50--300~keV band of
 hard events with observed 
fluence ratio $S(2-30~{\rm keV})/S(30-400~{\rm keV}) < 10^{-0.5}$.
We calculate the {\scriptsize ON} time duration as the time during which
the emission is larger than 10\% of the peak flux.
The subjet distribution is given by the power-law form.
The source redshifts are varied according to the cosmic star formation
rate.
}
\label{dgoP}
\end{figure}

\section{Discussion}
\label{sec:discuss}

We have investigated the $T_{90}$ duration distribution of GRBs under the
unified model of short and long GRBs proposed by \citet{yin04b},
and found that the model can reproduce the bimodal
distribution observed by BATSE.
In our model,
the crucial parameter is the multiplicity ($n_s$) of the subjets in the
direction of the observer.
The duration of an $n_s=1$ burst is determined by the angular
spreading time of one subjet emission, while
that of an $n_s \geq 2$ burst is determined by the time interval between
the observed first pulse and the last one.
These two different time scales naturally lead a division of the burst
$T_{90}$ durations into the short and long ones. 
We also performed a similar calculation for a Gaussian distribution,
$n(\vartheta,\varphi) = n_c
\exp\left[-(\vartheta/\vartheta_c)^2/2\right]$,
and found that the $T_{90}$ duration distribution is bimodal
in the same way as for the power-law subjet model.

\begin{figure}
\plotone{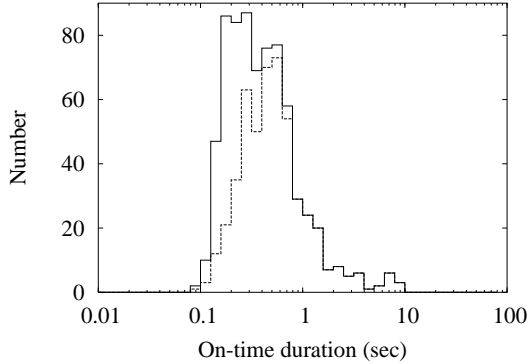}
\caption{
Same as Fig.~\ref{dgoP}, but 
the intrinsic luminosity of each subjet is assumed to be
$A^\j = A_{0} (= {\rm constant})$ for $\vartheta < 0.15$~rad and
$A^\j = A_{0}(\vartheta/0.15)^{-6}$ for
$0.15~{\rm rad}<\vartheta<\vartheta_b$,
where $A_{0}$ is in arbitrary unit.
Then we only take the events
with peak flux larger than $3 \times 10^{-4}$ of the maximum peak flux 
that has appeared in the calculation.
The dashed line represents $n_s \geq 2$ events. 
}
\label{dgotP}
\end{figure}

Let us make another comparison of our model with BATSE data. 
\citet{mitro98} have computed the distribution of the observed pulse
number (denoted by $n_p$ in their paper) and found that it is unimodal.
If the $n_p$ distribution were compared with the $n_s$ distribution, 
our model might be compatible with the observations, 
although some long GRBs are identified as $n_p=1$ events.
They also derive the distribution of the {\scriptsize ON} time
duration---defined as the time during which the emission is larger than
40\% of the peak flux---and found it bimodal.
Furthermore, they argue that the mean pulse widths of short and long GRBs
are different.
On the other hand, we computes the {\scriptsize ON} time duration
distribution in the context of our theoretical model and found it
unimodal (see Fig.~\ref{dgoP}), which is expected
since the pulse widths are almost the same. 
However, there are several observational implications that the
distances to short GRBs detected with BATSE are smaller than those of
long GRBs \citep[e.g.,][]{tava98,ggc04}, 
although this is controversial.
Then the observed pulse widths for short and long GRBs
might be different because of the redshift factor.
To give an example, let us assume that
the intrinsic luminosity of each subjet in the core region 
of the whole jet is larger than that in the periphery of the whole jet
and count only the GRB events with peak flux larger than
$3 \times 10^{-4}$ of the maximum peak flux in our simulation.
The result is shown in Figure~\ref{dgotP}, in which we find that
the effect of the peak flux cut off 
contributes to the bimodality of the {\scriptsize ON} time duration
distribution.

\begin{figure}
\plotone{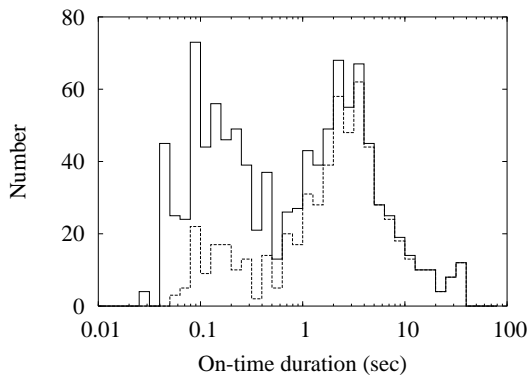}
\caption{
Same as Fig.~\ref{dgoP}, but
the emission radius $r^\j$ of each subjet is assumed to be
$r^\j = 3 \times 10^{14}$~cm for $\vartheta < 0.15$~rad and
$r^\j = 3 \times 10^{14} (\vartheta/0.15)^{-6}$~cm for
$0.15~{\rm rad}<\vartheta<\vartheta_b$.
The source redshifts are fixed as $z=1$.
The solid line represents all the events, while
the dashed line represents $n_s \geq 2$ events.
}
\label{dgom}
\end{figure}

At present, the observationally inferred  bimodality of {\scriptsize ON} time
duration is not explained in our current model, in which
all the subjets have the same intrinsic luminosity, the same opening 
half-angle, the same gamma factor, the same emission radius,
and so on. This is an extreme modeling for simple calculation.  
In reality, they may depend on the off-axis angle in the whole jet;
so may the pulse widths.
Furthermore,  
\citet{nakar02} investigated the pulse widths of GRBs using
2~ms time resolution and
report that short GRBs also consist of several pulses.
This can be incorporated into our model by assuming that
a subjet radiates successive emissions rather than 
one instantaneous emission.
Then the pulse width with 64~ms resolution \citep[which is used
in][]{mitro98} will be determined by the active time of the subjet.
If the pulse widths from the subjets in the central part
are larger than those in the periphery,
the bimodality of {\scriptsize ON} time duration distribution can be
explained.  
For example, we assume that
the emission radius $r^\j$ is larger for the core region than
for the periphery.
Figure~\ref{dgom} is the result, which shows the bimodal-like distribution.
Therefore, 
as we show in two examples (Figs.~\ref{dgotP} and \ref{dgom}),
some modifications of our model contribute to
the bimodality of the {\scriptsize ON} time duration, so that the
current observed {\scriptsize ON} time duration distribution is not
inconsistent with our model. 
We hope that in the future more sophisticated 
modeling will reproduce the observed {\scriptsize ON} time duration
distribution.

It has commonly been said that
the observed bimodal distribution of the $T_{90}$ durations of BATSE bursts
shows the different origins of short and long GRBs.
However, the bimodal distribution is also available as a natural
consequence of
our unified model of short and long GRBs.
The clear prediction of our unified model is that
short GRBs are associated
with energetic supernovae (SNe), since the association of long duration
GRBs with SNe is strongly suggested
\citep{galama1998,stanek2003,hjorth03,dellavalle2003}. 
Indeed, one of the short GRBs shows
possible association with a SN \citep{germany00}.
Even if the SNe are not identified with short GRBs because of
some observational reasons, we predict that the spatial distribution of
short GRBs in host galaxies should be similar to that of the long GRBs.
Another prediction is that short GRBs have the same total kinetic
energies as long GRBs, which might be confirmed by radio calorimetry
\citep{berger}.

\acknowledgements
We are grateful to the referee D.~Lazzati for instructive comments.
We would like to thank T.~Piran for useful discussions.
This work was supported in part by
a Grant-in-Aid for the 21st Century COE
``Center for Diversity and Universality in Physics''
and also by Grants-in-Aid for Scientific Research
of the Japanese Ministry of Education, Culture, Sports, Science,
and Technology 05008 (R.~Y.), 14047212 (T.~N.), and 14204024 (T.~N.).

\appendix
\section{Analytical Estimate of the intrinsic $T_{90}$ distribution of the
Long Bursts}
\label{sec:app}
In this Appendix we derive the analytical distribution function of
the $T_{90}$ durations of the long GRBs when all sources are assumed to be at
$z=1$.
At first we consider for a given $n_s(\geq 2)$.
Each subjet causes one pulse, whose shape is a
$\delta$-function for simplicity.
In the present case the arrival time of the pulse from each subjet
is random in the range $0<T_{\obs}<T_{\rm dur}$.
For a given $T_{90}$, the first pulse is
required to arrive within $T_{\rm dur}-T_{90}$.
The arrival time of the last pulse is determined as the time
$T_{90}$ after the first pulse.
The rest of the pulses are required to arrive within the range of $T_{90}$.
Thus, the probability function of $T_{90}$ for a fixed $n_s$ is
approximately given by
\begin{equation}
P_{n_s}(T_{90})dT_{90} = n_s(n_{s}-1)\frac{T_{\rm dur}-T_{90}}{T_{\rm dur}}
\left(\frac{T_{90}}{T_{\rm dur}}\right)^{n_{s}-2}
\frac{dT_{90}}{T_{\rm dur}}~~.
\end{equation}
For the power-law angular distribution of the subjets
the distribution function of $n_s$ is proportional to ${n_s}^{-2}$,
so that we get
\begin{equation}
P(T_{90})dT_{90} \propto \sum_{n_s=2}^{\infty} {n_s}^{-2} P_{n_s}(T_{90})
dT_{90}= \frac{(T_{90}/T_{\rm dur})+[1-(T_{90}/T_{\rm dur})]
\log[1-(T_{90}/T_{\rm dur})]}{T_{90}/T_{\rm dur}}\frac{dT_{90}}{T_{90}}~.
\label{eq:appP}
\end{equation}
The distribution function of $n_s$ for the Gaussian angular distribution
of the subjets can be obtained in a similar way.


\end{document}